\documentclass[journal, compsoc]{IEEEtran} % compsoc gives us arabic numbering
\usepackage{graphicx}
\usepackage[utf8]{inputenc}
\usepackage{hyperref} % provides url styling
\usepackage{amssymb} % provides number set notation
\usepackage{amsmath} % provides eqref command
\usepackage{caption}
\usepackage{subcaption}
\usepackage{natbib}
\usepackage{siunitx} % provides the ang command typesetting angles
\usepackage{tabularx}
\usepackage{multirow} % provides multirow environment for tables

\setcitestyle{authoryear}

\makeatletter
\def\@seccntformatinl#1{\csname the#1dis\endcsname\hskip 1em\relax}
\makeatother

\begin{document}
%
% paper title
\title{SimTreeLS: Simulating aerial and terrestrial laser scans of trees}

\author{Fred~Westling*,
    Mitch~Bryson,
    James~Underwood% <-this % stops a space
    \thanks{F. Westling and M. Bryson are with the University of Sydney}% <-this % stops a space
    \thanks{J. Underwood is with Green Atlas}% <-this % stops a space
    \thanks{*Correspondence: f.westling@acfr.usyd.edu.au}
}

% make the title area
\maketitle

\begin{abstract}
    There are numerous emerging applications for digitizing trees using terrestrial and aerial laser scanning, particularly in the fields of agriculture and forestry.
    Interpretation of LiDAR point clouds is increasingly relying on data-driven methods (such as supervised machine learning) that rely on large quantities of hand-labelled data. As this data is potentially expensive to capture, and difficult to clearly visualise and label manually, a means of supplementing real LiDAR scans with simulated data is becoming a necessary step in realising the potential of these methods.
    We present an open source tool, SimTreeLS (Simulated Tree Laser Scans), for generating point clouds which simulate scanning with user-defined sensor, trajectory, tree shape and layout parameters.
    Upon simulation, material classification is kept in a pointwise fashion so leaf and woody matter are perfectly known, and unique identifiers separate individual trees, foregoing post-simulation labelling.
    This allows for an endless supply of procedurally generated data with similar characteristics to real LiDAR captures, which can then be used for development of data processing techniques or training of machine learning algorithms.
    To validate our method, we compare the characteristics of a simulated scan with a real scan using similar trees and the same sensor and trajectory parameters.
    Results suggest the simulated data is significantly more similar to real data than a sample-based control.
    We also demonstrate application of SimTreeLS on contexts beyond the real data available, simulating scans of new tree shapes, new trajectories and new layouts, with results presenting well.
    SimTreeLS is available as an open source resource built on publicly available libraries.
\end{abstract}


\IEEEpeerreviewmaketitle

\section{Introduction}
\label{sec:introduction}

% Why is LiDAR scanning of trees useful?
LiDAR scanning is a useful tool for reality capture in various industries.
In agriculture, \cite{rosell2012review} showed that LiDAR is considered a useful tool for rapidly capturing geometric properties of trees.
\cite{wu2020suitability} confirmed the suitability of LiDAR from both a terrestrial and aerial context for analysing tree crop structures.
Scanning allows extraction of tree parameters which are useful for growth analysis, including woody matter detection (\cite{vicari2019leaf,su2019extracting,westling2020graph}), porosity~(\cite{pfeiffer2018mechatronic}), and leaf area density or distribution~(\cite{beland2011estimating,beland2014model,sanz2018lidar}.
Beyond simple tree parameters, LiDAR scanning also enables detailed investigations on real trees, like orchard mapping (\cite{underwood2016mapping,reiser2018iterative}), analysis of the light environment of a tree (\cite{westling2018light}), or historical yield analysis (\cite{colacco2019spatial2}).  \cite{gene2019fruit} was even able to detect fruit using intensity returns from a terrestrial LiDAR scanner, with comparable results and several advantages over vision systems.
Similar applications are found in forestry, where tree geometric analysis is of interest.
Tree parameters like woody matter detection (\cite{Ma2016,Ma2016a}) and Leaf Area Density (\cite{VanderZande2011})is also of interest here, as well as further interest areas including robot navigation (\cite{Lalonde2006}) and forest inventory (\cite{bauwens2016forest}).
Terrestrial laser scanning is used extensively in commercial forests for plot-level inventory (\cite{liang2016terrestrial}), which is then used for validation and training for inventory over large scales based on aerial LiDAR capture (\cite{Kato2009,wang2020characterizing,cao2019comparison,almeida2019monitoring}).  \cite{larue2020compatibility} demonstrates that aerial LiDAR captures slightly less than the terrestrial equivalent, but is well suited to macro-scales.
In both industries however, LiDAR capture involves time-intensive scanning operations, and captured data can be difficult to process.

% Who else has done virtual trees?
There is some value and interest for analysing trees in silico to achieve perfect digitization, generate large datasets, and perform physically challenging or destructive operations.
\cite{yang2016canopy} and \cite{arikapudi2015orchard} digitized trees thoroughly for light interception analysis and geometric modelling respectively.
Others have generated virtual trees using algorithmic growth (for example Functional-Structural Plant Modelling presented by \cite{White2012,White2016}) which allows for study of how a tree will develop using different pruning or growth decisions.
Sometimes the approach is to generate computer models of particular trees, for instance \cite{da2014light} and \cite{tang2015light} performed light interception efficiency analyses using computer models of apple and peach trees respectively, while \cite{tang2019optimal} generated virtual loquat trees to design an optimal plant canopy shape.
Beyond simple computer modelling, \cite{tao2015geometric} used Physically Based Ray Tracing to simulate terrestrial Lidar scans, based on an approach previously used by \cite{Cote2009}.
Some of these methods generate perfect data while others approximate real scanning, and both approaches have value in different application areas.

% Rise of deep learning
Recently, there has been a growing interest in deep learning on point clouds, as reviewed by \cite{guo2019deep}.
For general point cloud applications, a variety of approaches have been developed, from the multi-view convolutional neural networks presented by \cite{su2015multi} to dense contextual networks \cite{liu2019densepoint}.  However, many methods primarily operate on small or perfectly sampled point clouds (\cite{wu20153d,qi2017pointnet}), and applications on tree scans are typically neither.
\cite{kumar2019development} identified trees as distinct from other object types in point clouds captured by mobile laser scanning with a total accuracy of 95.2\%.
In forestry, \cite{windrim2018forest} and \cite{xi2018filtering} perform tree classification using fully connected 3D CNNs.
In agriculture, \cite{majeed2020deep} used deep learning to segment plant matter from its supporting trellis; however, typically deep learning here is applied to imagery rather than LiDAR (\cite{bargoti2015pipeline,apolo2020cloud}).
A significant obstacle to deep learning on large-scale point clouds is the difficulty in acquiring large quantities of data which is labelled by human experts.
Modern deep learning architectures contain potentially hundreds of thousands to millions of trainable parameters in order to make accurate and robust inference.
These models demand the use of tens to hundreds of thousands of training examples to realise the potential of this complexity, and providing all of these training labels manually on real data can quickly become infeasible.
LiDAR scanning must typically decide on a trade-off between quality and capture speed, with faster captures typically being more sparse or occluded and thus harder to label (\cite{westling2020graph}).

% Why is simulation good? Not tree-specific
Simulation of realistic data is a viable candidate for solving the labelling issue.
Many deep learning applications, in particular standard datasets and early point cloud works, use point-sampled meshes to generate point clouds on which to learn (for example, \cite{wu20153d} and \cite{qi2017pointnet}), though this data does not realistically estimate the effect of laser scanning an object.
Developing this further, \cite{wang2019automatic} and \cite{goodin2019training} have shown that neural networks can be trained comparably well using simulated LiDAR data.
Generally in machine learning, transfer learning allows learners to minimise the required dataset by training on a similar set first, though there are complexities (\cite{zhuang2020comprehensive}).
\cite{nezafat2019transfer} were able to transfer features learned by a pre-trained model to a different data setting, using images generated by projection of LiDAR data.
This suggests that realistic simulated data which is perfectly labelled could be used to pre-train models, significantly reducing the need for manual labelling.
This can also be applied to changes in setting, for instance comparing aerial to terrestrial LiDAR.

\begin{figure}[ht]
    \includegraphics[width=\columnwidth]{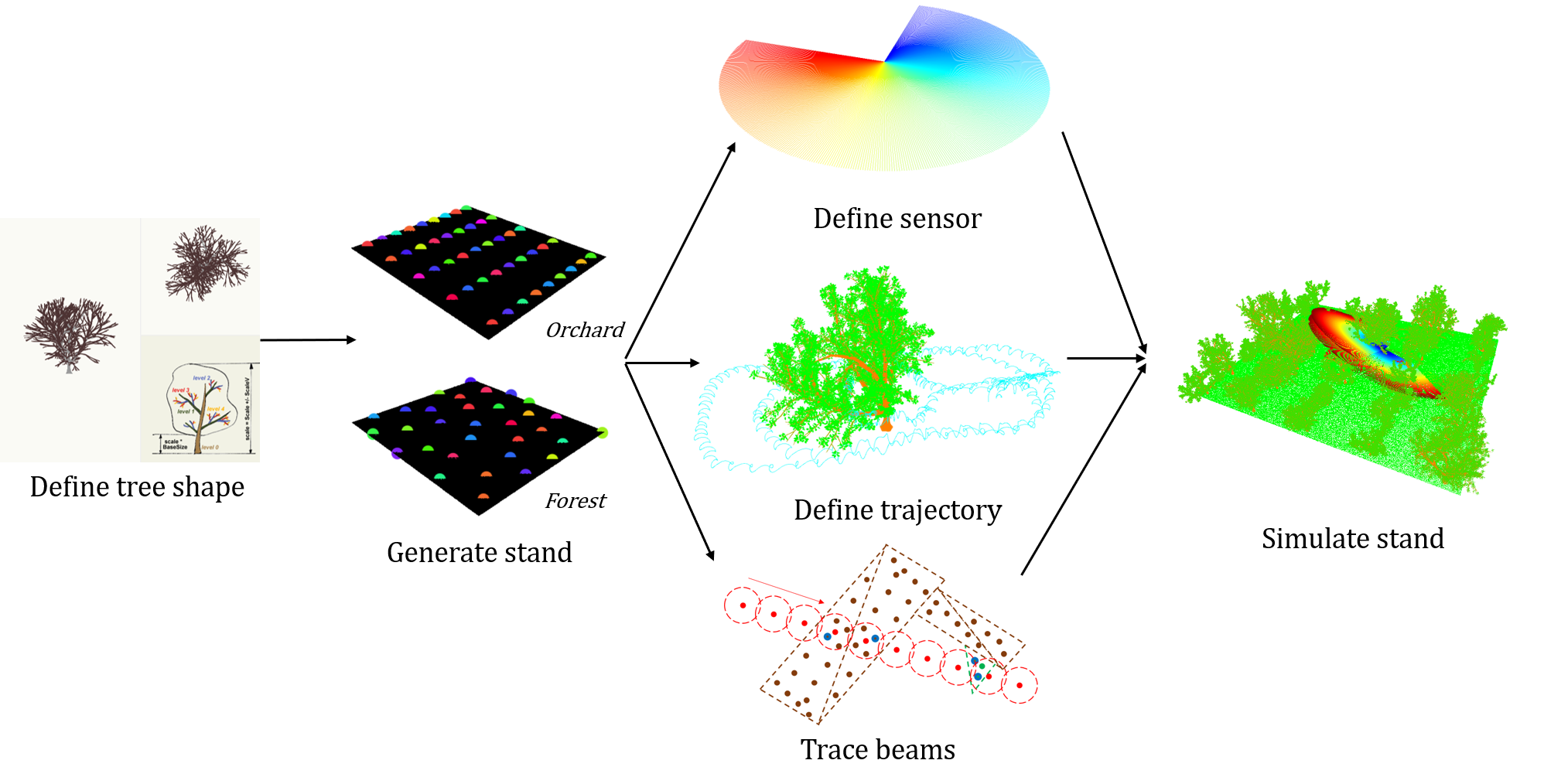}
    \caption{Graphical overview of how SimTreeLS works}
    \label{fig:intro-abstract}
\end{figure}

% What are we doing and why?
We present a software tool, SimTreeLS, to generate simulated LiDAR scans with realistic sensor parameters, trajectories and results.
Other tools with similar aims have been presented like SIMLIDAR by \cite{mendez2012simlidar}, though not at the same scale or level of detail.
Our approach, SimTreeLS, is specifically designed to create simulated scans of trees, for development of applications in agriculture and forestry, supports a wide variety of tree shapes and sensor types, and can be used to simulate ground-based, handheld and aerial mobile LiDAR.
We present how SimTreeLS works and experimental results showing its viability as a simulated source of LiDAR data.
An overview of our method is presented in Figure~\ref{fig:intro-abstract}.

\section{Method}
\label{sec:method}
% Introductory paragraph
SimTreeLS is designed to be extensible to a range of situations.
In this section, we describe how the system is set up, from defining tree shape and organisation through to simulating the scanning process itself.
We then explain the validation experiments with which we demonstrate the suitability and capabilities of SimTreeLS for use as a data generation tool.
The core components of SimTreeLS are the open source libraries Comma and Snark (\cite{acfrcomma}).
SimTreeLS is available at https://github.com/fwestling/SimTreeLS.

% ========= SimTreeLS ========= %
\subsection{Simulating scans}
% Tree definitions
\subsubsection{Defining trees}
\begin{figure}[ht]
    \includegraphics[width=\columnwidth]{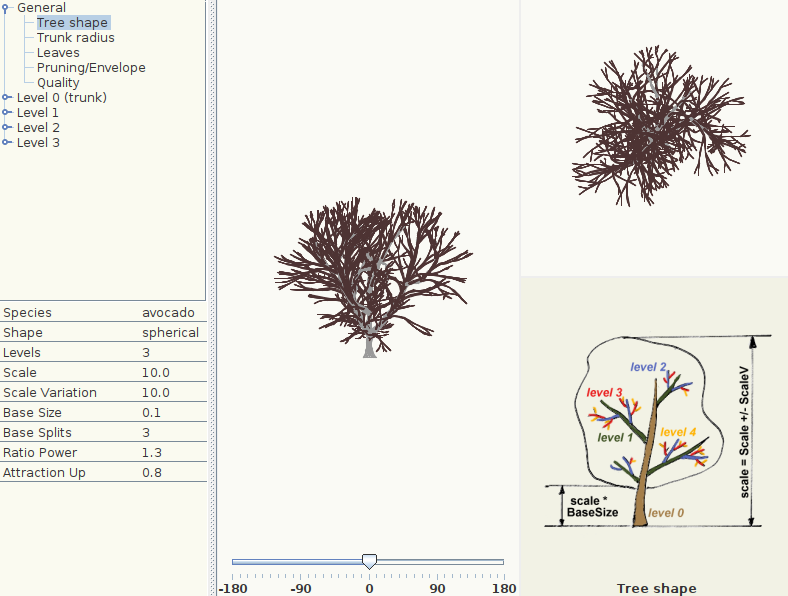}
    \caption{Screenshot of Arbaro tree generation software with a customised tree definition (\cite{diestel2003arbaro})}
    \label{fig:method-arbaro}
\end{figure}

To generate tree model objects, we use the Arbaro software made by \cite{diestel2003arbaro} and based on the work of \cite{weber1995creation}.
Arbaro provides a number of predefined realistic tree definitions, and allows the user to create new definitions to suit the tree shapes in which they are interested, using the interface shown in Figure~\ref{fig:method-arbaro}.
The tree definitions are encoded in XML format, which is used as an input for SimTreeLS.
We run the Arbaro command line interface using a tree definition to generate an individual tree as a 3d mesh in OBJ format.
By providing varying seeds for a random generator, different individuals can be produced using the same definition, and we use the seed of a particular individual tree as its ID for later processing.

When generating a dataset, SimTreeLS calculates how many trees are required and generates the necessary OBJ files, then performs mesh sampling to turn the tree meshes into very high density point clouds with no noise.
This is required for our sampling method used later in the process; although the mesh file could potentially be used directly, our method (described below) operates on point clouds.
The sampling density is chosen to be high enough that the distance between points is smaller than the search radius for a single sensor beam, such that if a beam passes through an element of the mesh it should detect the sampled points at that location.

Simulating the trees like this provides perfectly and completely labelled data, as every point in the point cloud is marked according to which level of branch it represents, from Level 0 (trunk) to Level 3 (leaf).

% Constructing stands
\subsubsection{Organising stands}
\begin{figure}[ht]
    \centering
    \begin{subfigure}[t]{\columnwidth}
        \centering
        \includegraphics[width=\textwidth]{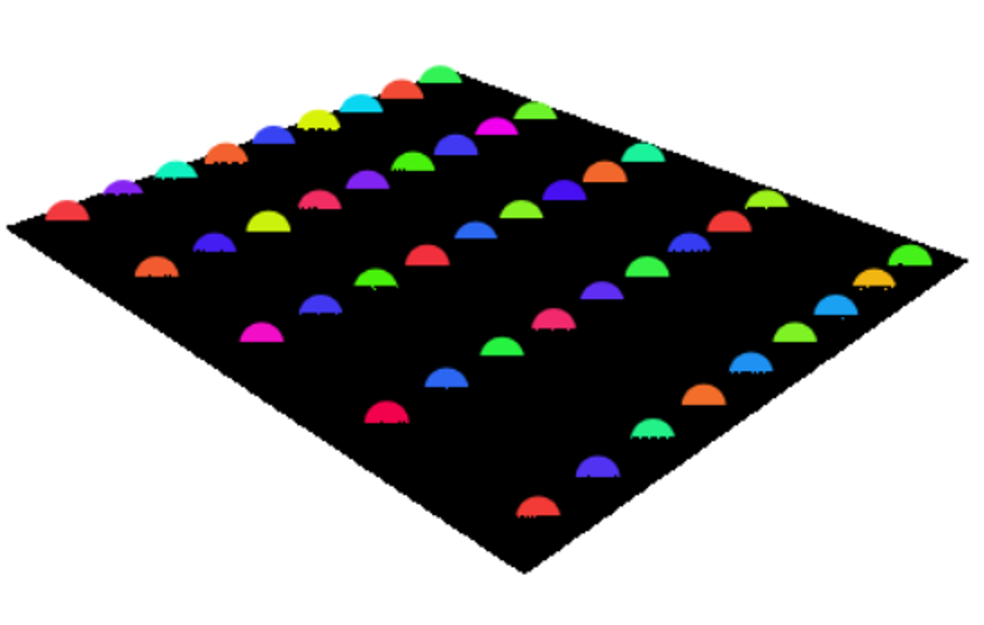}
        \caption{Orchard}
    \end{subfigure}
    ~
    \begin{subfigure}[t]{\columnwidth}
        \centering
        \includegraphics[width=\textwidth]{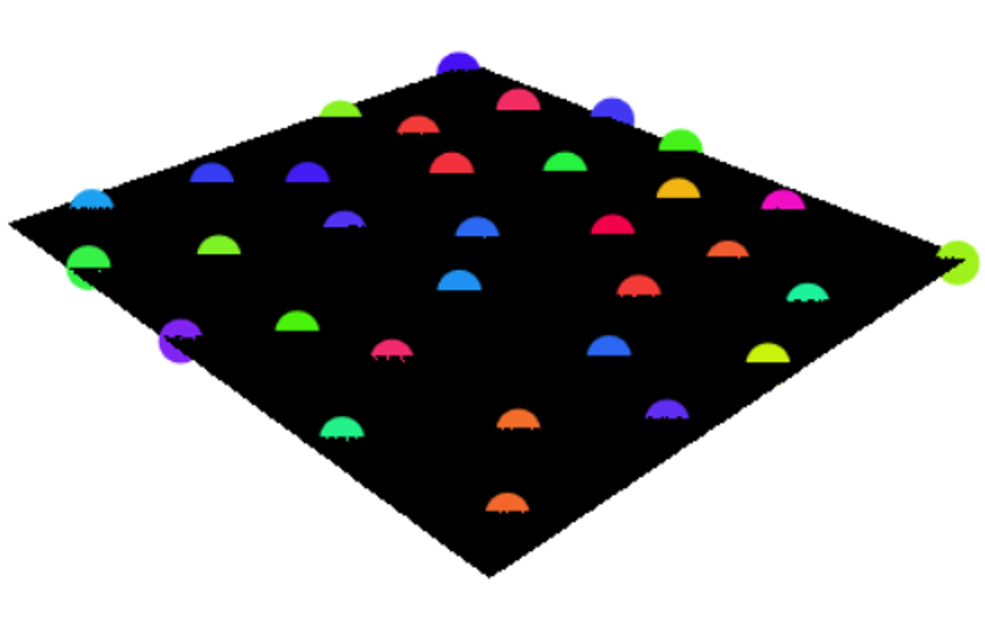}
        \caption{Forest}
    \end{subfigure}
    \caption{Different stand layouts for agriculture and forestry.  The orchard layout is procedural with row and tree spacing as user-defined parameters, while the forest layout is non-deterministically generated while enforcing a minimum spacing between trees.  Each coloured point shown is selected as the trunk location of a simulated tree to be added later.  The spacing of the orchard trees is 4m (tree) / 10m (row) for easy of visibility, though for our experiments we use 6m / 10m.  The spacing for the forest trees in this example has been set to a minimum of 6m.}
    \label{fig:method-stands}
\end{figure}

SimTreeLS is designed primarily for use in agriculture and forestry, where trees are typically processed at large scale rather than individually.
Therefore, we included the ability to simulate collections of trees that we call "stands" in two pre-defined formats, shown in Figure~\ref{fig:method-stands}.
The "orchard" format organises trees in straight rows, with user-defined values for tree spacing and row spacing.
The rows are by default north-aligned, but can easily be rotated to any angles, which has allowed orientation experiments presented by \cite{westling2020replacing}.
The "forest" format arranges trees randomly, with a user-defined minimum distance between them.

% Sensors and trajectories
\subsubsection{Sensors and trajectories}
\begin{figure}[ht]
    \centering
    \begin{subfigure}[t]{\columnwidth}
        \centering
        \includegraphics[width=\textwidth]{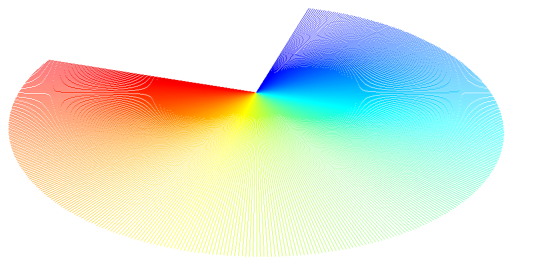}
        \caption{Single-beam LiDAR}
    \end{subfigure}
    ~
    \begin{subfigure}[t]{\columnwidth}
        \centering
        \includegraphics[width=\textwidth]{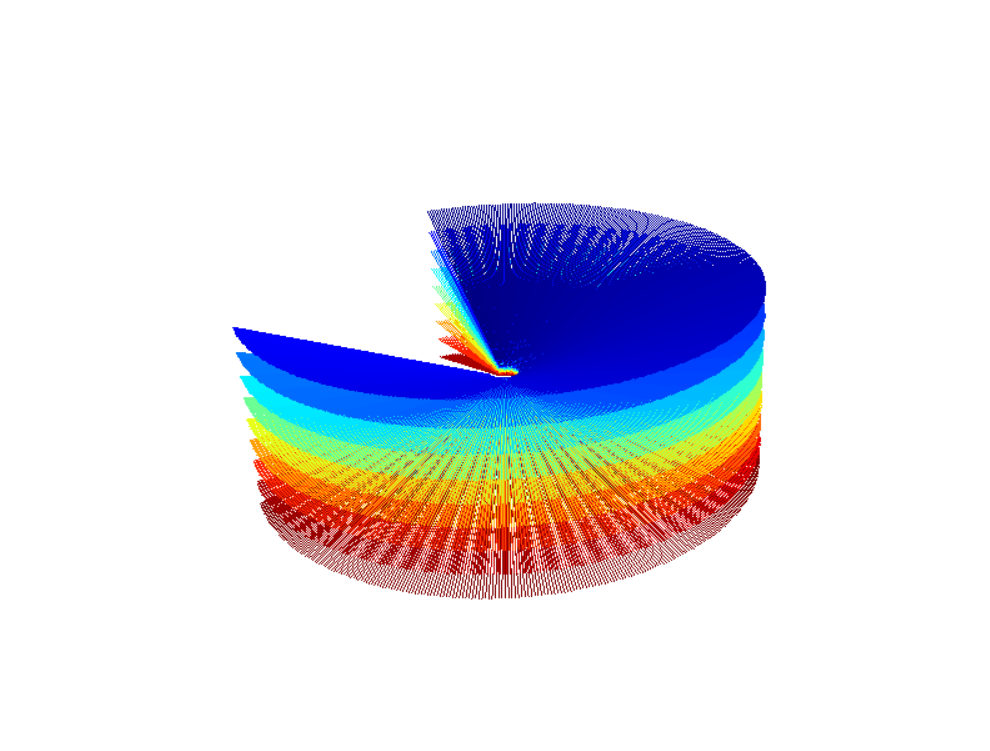}
        \caption{9-beam LiDAR}
    \end{subfigure}
    \caption{Example sensor shape definitions for two different LiDAR sensors.  The sensor has been simulated with $270\deg$ range, $0.675\deg$ angular resolution and 15m range, and presented as a single-plane LiDAR as well as a 9-beam LiDAR with a vertical range of $30\deg$.  Each beam has a unique identifier related to its angular position, here coloured from blue to red.}
    %Hokuyo UTM-30LX

    \label{fig:method-sensorshape}
\end{figure}

\begin{figure}[ht]
    \centering
    \begin{subfigure}[t]{\columnwidth}
        \centering
        \includegraphics[width=\textwidth]{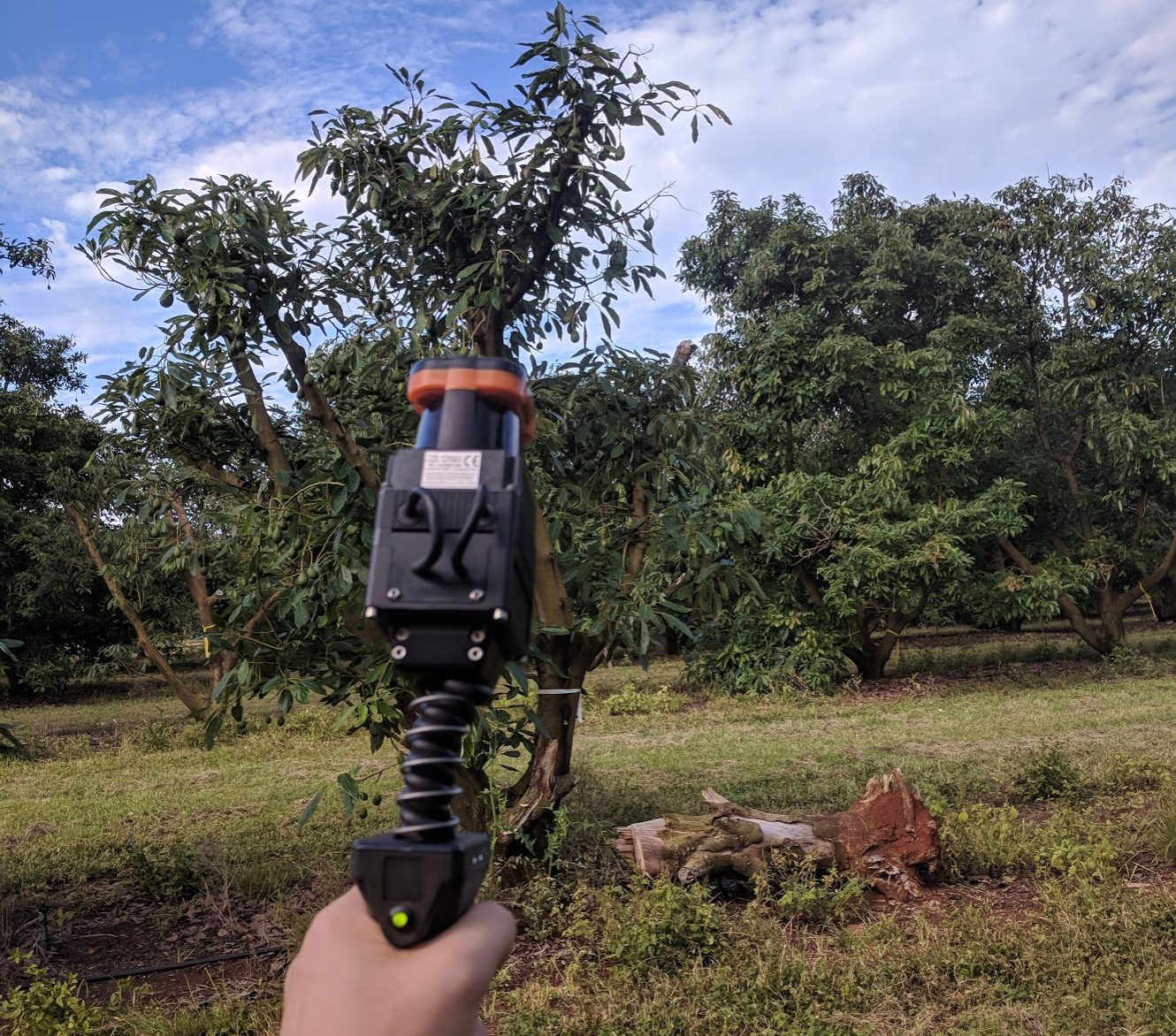}
        \caption{Sensor used to generate trajectory}
    \end{subfigure}
    ~
    \begin{subfigure}[t]{\columnwidth}
        \centering
        \includegraphics[width=\textwidth]{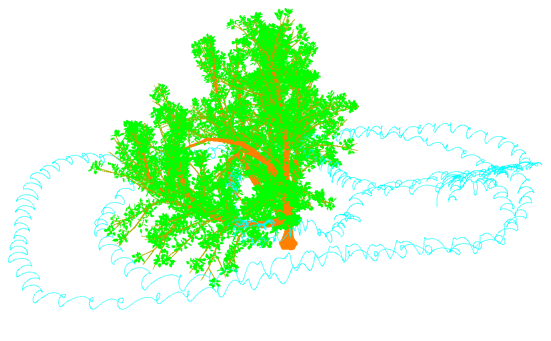}
        \caption{Trajectory in silico}
    \end{subfigure}
    \caption{Example sensor trajectory based on the trajectory of a GeoSLAM Zeb1.  The trajectory is here overlaid on a virtual tree with trunk located at (0,0,0).  The path followed is a double closed-loop trajectory employing a wide circle for context and a close circle for occlusion minimisation.  The sensor generates a 3D scan by oscillating the 2D sensor, causing the particular shape seen in the trajectory.}
    \label{fig:method-trajectory}
\end{figure}

To simulate realistic LiDAR scans, we define a LiDAR sensor shape and a trajectory.
The sensor shape is created as a point cloud such that it is not limited to simple planes, but can easily be extended to other definitions like composites (e.g. multi-plane LiDAR) or spherical (e.g. rapidly rotating LiDAR).
For any sensor definition, the points should be organised as "scan lines" which represent the path of the laser at any given time.
The distance between points within a scan line is determined by the distance accuracy and angular resolution of the specific sensor being simulated, and each individual scan line is assigned a unique identifier to control occlusion.
Figure~\ref{fig:method-sensorshape} shows some example LiDAR definitions.
This format allows us to use point cloud comparison tools provided by Snark, as well as allowing multi-plane sensors like the Velodyne Puck to be simulated as easily as a 2D sensor, as is shown in the figure.
Trajectories are simply defined as a list of points with associated quaternions to define the pose of the LiDAR at each time step.
Figure~\ref{fig:method-trajectory} shows an example trajectory based on a scan using a handheld sensor (GeoSLAM Zeb1, \cite{bosse2012zebedee}).  This trajectory is a loop around a central tree designed to maximise the information captured on that tree, with little concern for surroundings.
This trajectory is useful for studying SimTreeLS due to its complexity and because it was designed to minimise occlusion, but simpler trajectories can also be used.

Traditional approaches to ray tracing such as that presented by \cite{li2020simulation} define rays as lines and keep the mesh being scanned as a list of shapes, whereas we represent both the sensor and the mesh as a point cloud.
While this leads to larger memory requirements on the data, it enables optimisations and parallelisation using existing point cloud processing tools.
In particular, this data formulation allows the use of the Snark library, which is optimised for fast streaming processes on point cloud data.
Other commonly used optimisations for point clouds, for instance octrees, could further be applied to improve processing speeds.

% Simulating LiDAR scans
\subsubsection{Simulating LiDAR scans}
\begin{figure}[ht]
    \centering
    \begin{subfigure}[t]{\columnwidth}
        \centering
        \includegraphics[width=0.9\textwidth]{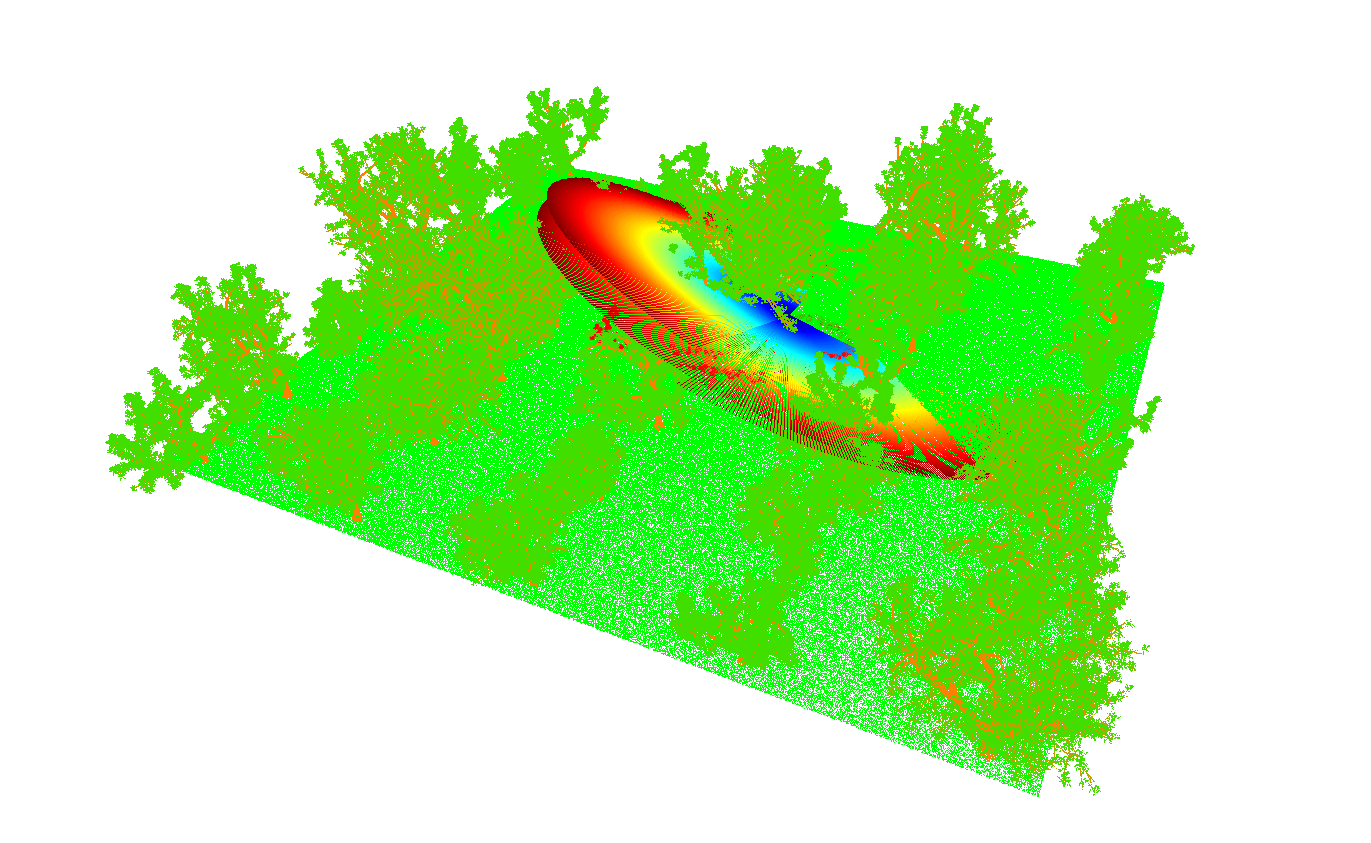}
        \caption{GeoSLAM Zeb1}
    \end{subfigure}
    ~
    \begin{subfigure}[t]{\columnwidth}
        \centering
        \includegraphics[width=0.9\textwidth]{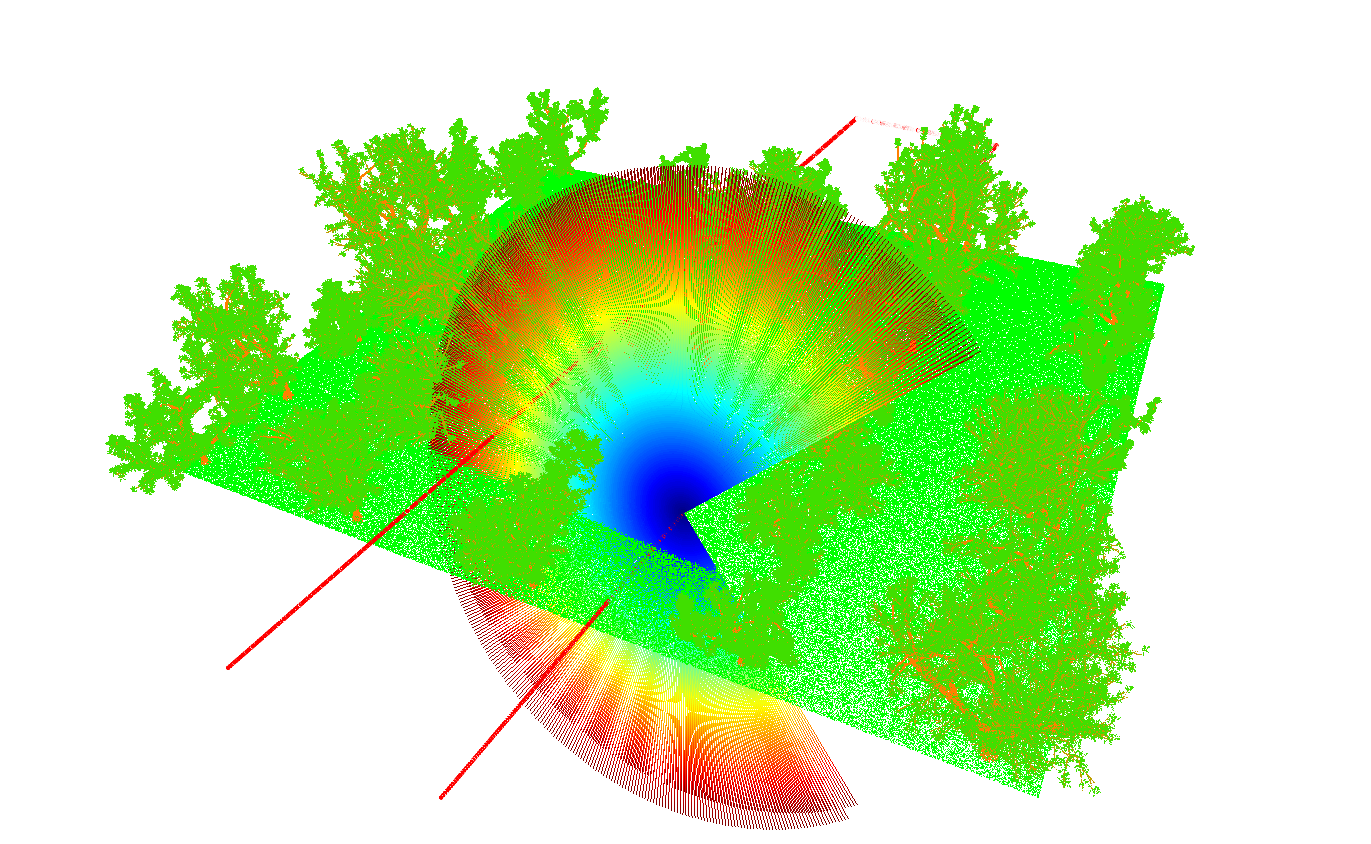}
        \caption{Mobile platform}
    \end{subfigure}
    \begin{subfigure}[t]{\columnwidth}
        \centering
        \includegraphics[width=0.9\textwidth]{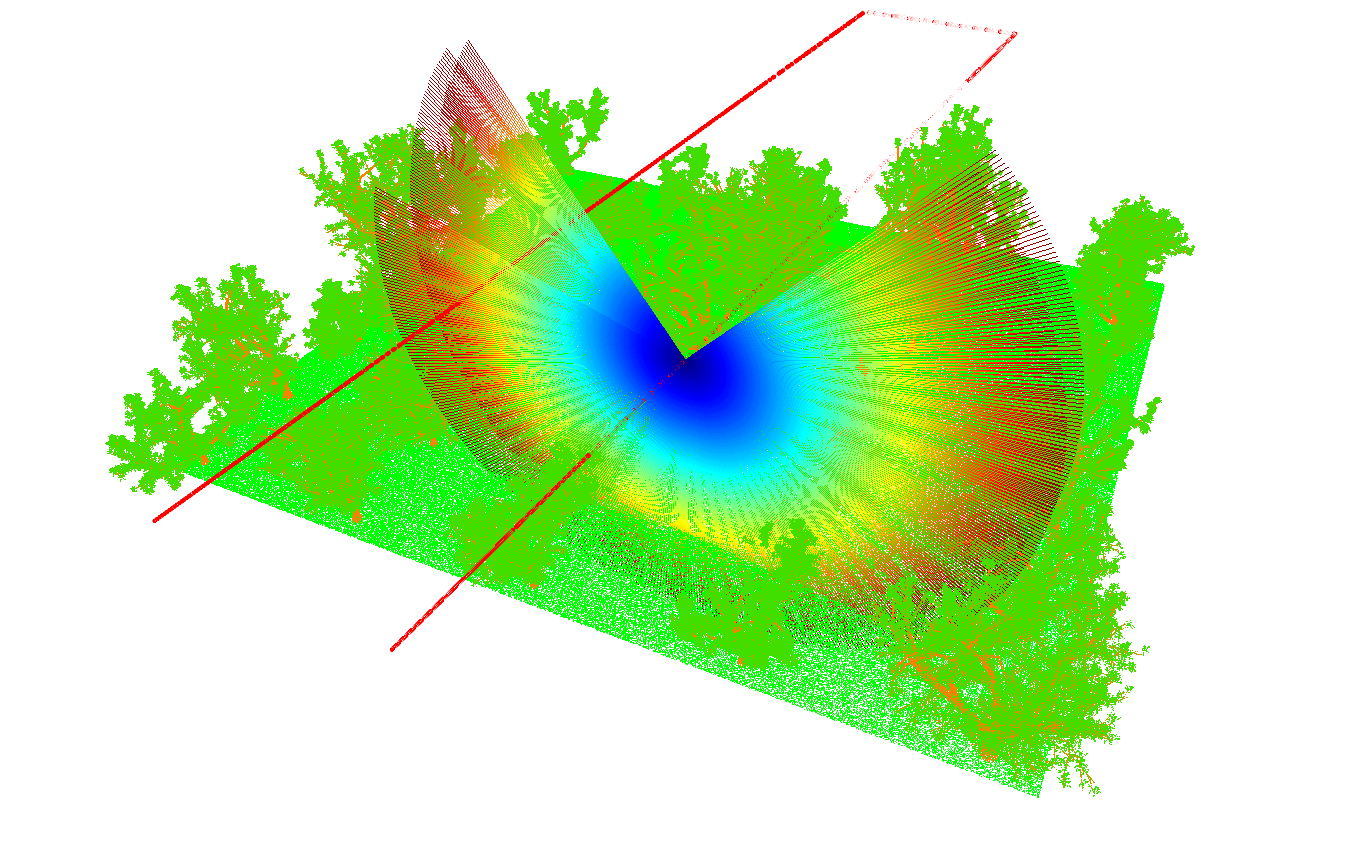}
        \caption{Aerial scan}
    \end{subfigure}
    \caption{Snapshot visualisation of scan simulation, with the sensor shape at a point in the trajectory overlaid on a stand of trees (orchard, 6m / 10m spacing).  Here we show three different trajectories, simulating handheld, mobile and aerial LiDAR.}
    \label{fig:method-demos}
\end{figure}

The process of simulating the LiDAR scan is relatively simple, since the stands, trajectory and sensor are defined.
The sensor shape is moved to each point in the trajectory in turn, and from each location, the scanning process is performed.
A snapshot of the sensor placement process is illustrated for three different trajectories in Figure~\ref{fig:method-demos}.

\begin{figure}[ht]
    \centering
    \includegraphics[width=\columnwidth]{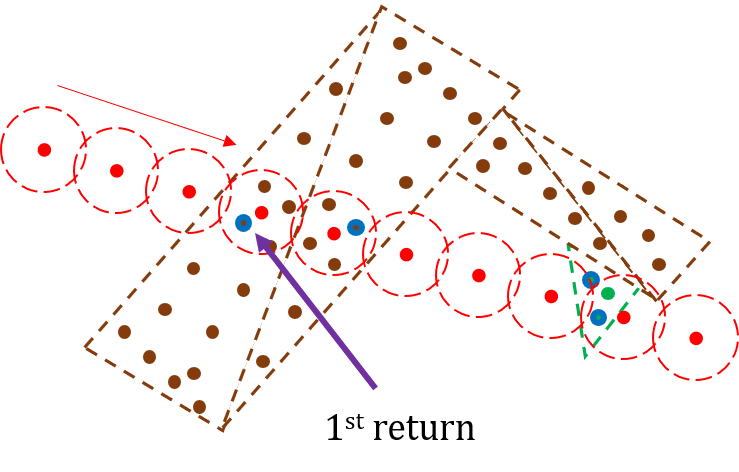}
    \caption{Illustration of scan simulation process for a single scan line in the high density point cloud.  Brown and green points represent the tree being scanned.  Red points represent the points in the scan line, with a dotted border to signify the search radius.  Points circled in blue are identified as candidate points detected by this scan line, and only the first one is returned, such that the green leaf is "occluded" from this scan line.}
    \label{fig:method-scanline}
\end{figure}

At each location, the points of the sensor shape are mapped to their nearest point in the high-density point cloud representing the stand of trees.
Each point in the scan plane is only matched to points within a certain radius, selected as the error of the sensor being simulated.
Since each scan line has a unique identifier, only match closest to the sensor origin for each scan line is kept, so that occlusion is handled correctly.
This process is illustrated in Figure~\ref{fig:method-scanline}.
Once a first-return point has been selected, gaussian noise with a user-defined mean is added to the coordinates of that point to simulate the effects of sensor noise and tree movement due to wind.

% ========= Experiments ========= %
\subsection{Validation Experiments}

Since the input tree objects are generated in silico, there is no option to collect ground truth data using real LiDAR.
However, our aim is to validate that the point clouds produced by SimTreeLS are similar in nature to those captured by real LiDAR, so we can compare metrics which are characteristic of LiDAR scans in these environments.

The metric we are demonstrating is the scan density: when scanning with a real LiDAR, the density of the point cloud is non-uniform and dependent on the distance of the object from the sensor, occlusion, and a variety of other influences.  We examined the average and deviation of the point cloud density as a function of height and distance from the origin.

To validate our scans, we compared a representative scan of avocado trees using a GeoSLAM Zeb1 handheld LiDAR with point cloud generated by SimTreeLS.
We also generated control point clouds by performing mesh sampling without any sensor simulation, at a sampling density which leads the final cloud to have a similar number of points as the simulated clouds.
This is the easiest way to turn the high density point cloud into simulated point cloud data, but does not incorporate any occlusion or sensor range limitations.
We aimed to show that SimTreeLS outputs are closer to real LiDAR scan data than those generated by this simpler approach.

Since the intent of these simulations is to generate usable data, we also test the applicability of the results to tree analysis operations.
Specifically, we used the graph-based trunk classification and individual segmentation process presented by \cite{westling2020graph} on large quantities of real data as well as virtual data, and performed a two-tailed unpaired T test to see whether the results are conceivably from similar distributions.

% ========= Applications ========= %
\subsection{Applications}

Simulated tree data has other applications beyond simply replacing real data.
In this section, we describe the experiments we performed to illustrate some of these applications.

One such application is the capability of SimTreeLS to simulate a range of scanning contexts, which can be used to estimate the value of a particular sensing modality.
Validation experiments described in the previous section were conducted using the "Zeb1" trajectory, which simulates the trajectory of the handheld LiDAR.  Other common trajectories for orchard scanning are terrestrial mobile platforms like the SHRIMP platform presented by \cite{underwood2016mapping} or scanning by aerial drone-mounted LiDAR such as \cite{almeida2019monitoring}.
We simulated scans of different trees scanned by simulated trajectories in these additional contexts to show the difference in density characteristics.

Another application made possible by simulated data is the analysis of occlusion, which is a significant reason that real scans may not show the full context of the scene.
To investigate this, we examined the occlusion profile of different scanning contexts as compared to a simply sampled control.
We estimated occlusion by mapping the point cloud from the simulated scan back to the original high-quality point cloud to identify occluded matter, within a small search radius.

\begin{figure}[ht]
    \centering
    \begin{subfigure}[t]{\columnwidth}
        \centering
        \includegraphics[width=0.52\textwidth]{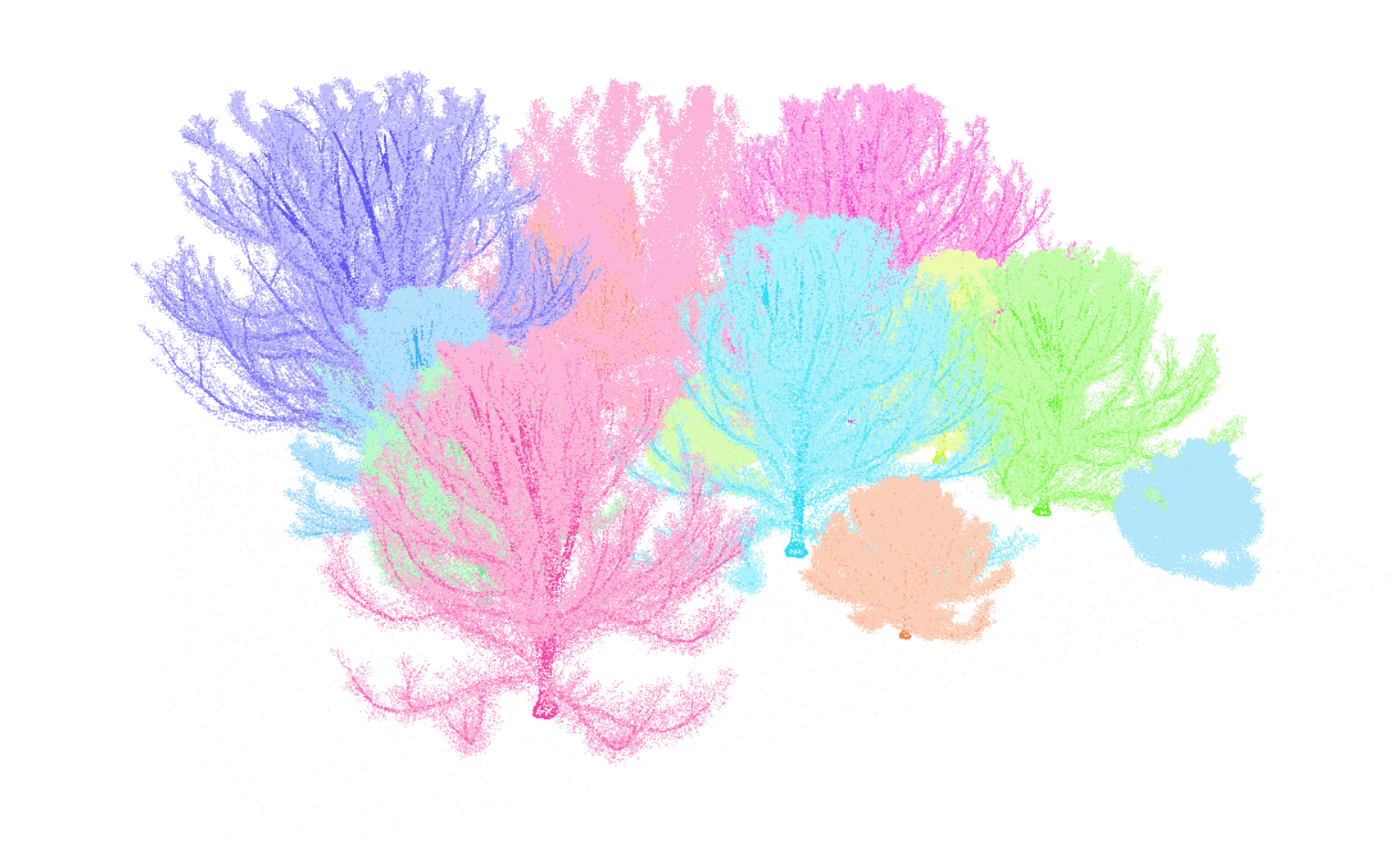}
        \caption{Macadamia orchard}
    \end{subfigure}
    ~
    \begin{subfigure}[t]{\columnwidth}
        \centering
        \includegraphics[width=0.52\textwidth]{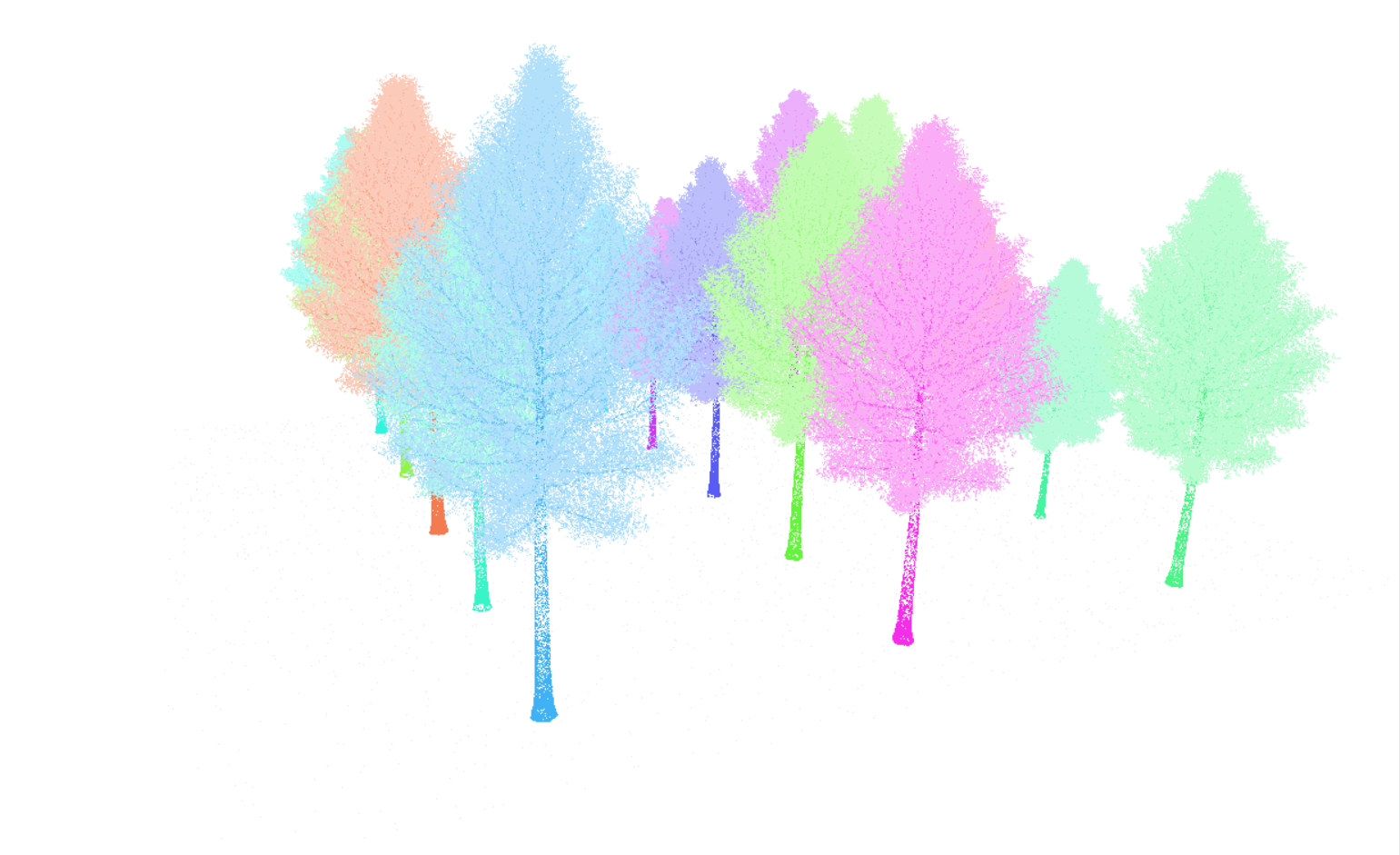}
        \caption{Aspen orchard}
    \end{subfigure}
    ~
    \begin{subfigure}[t]{\columnwidth}
        \centering
        \includegraphics[width=0.52\textwidth]{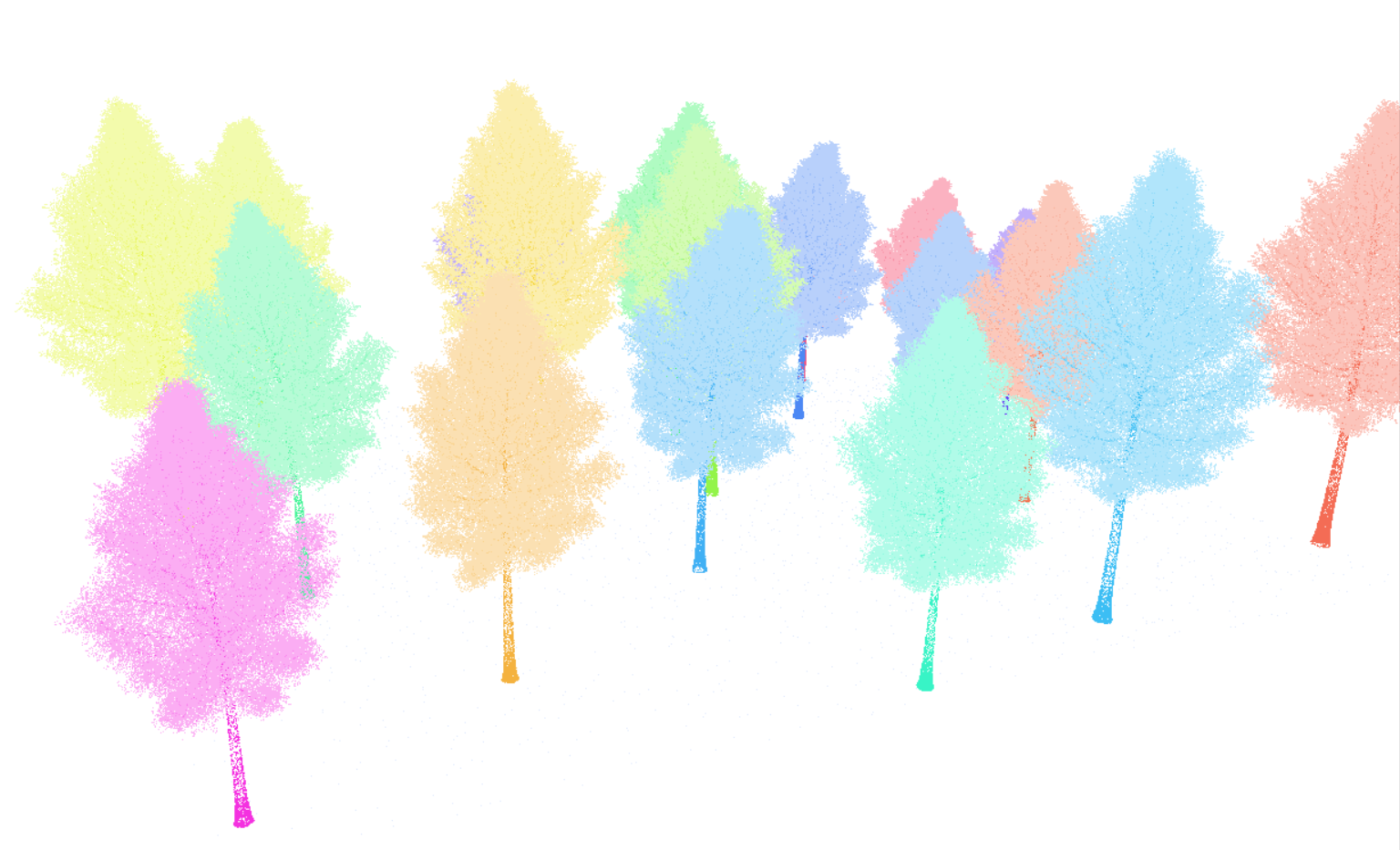}
        \caption{Aspen forest}
    \end{subfigure}
    \caption{Additional tree and stand definitions used in application experiments.}
    \label{fig:show-demos}
\end{figure}

We also performed the same analyses of density and occlusion using two additional tree definitions to show the flexibility of the system, as well as a forest-style stand, with a random distribution of trees at a minimum tree spacing of 6m. The additional datasets are visualised in Figure~\ref{fig:show-demos}.

\section{Results}
\label{sec:results}
% Introductory paragraph
In this section, we present the results of basic SimTreeLS operation.

\subsection{SimTreeLS outputs}

\begin{figure}[ht]
    \centering
    \begin{subfigure}[t]{\columnwidth}
        \centering
        \includegraphics[width=0.4\textwidth]{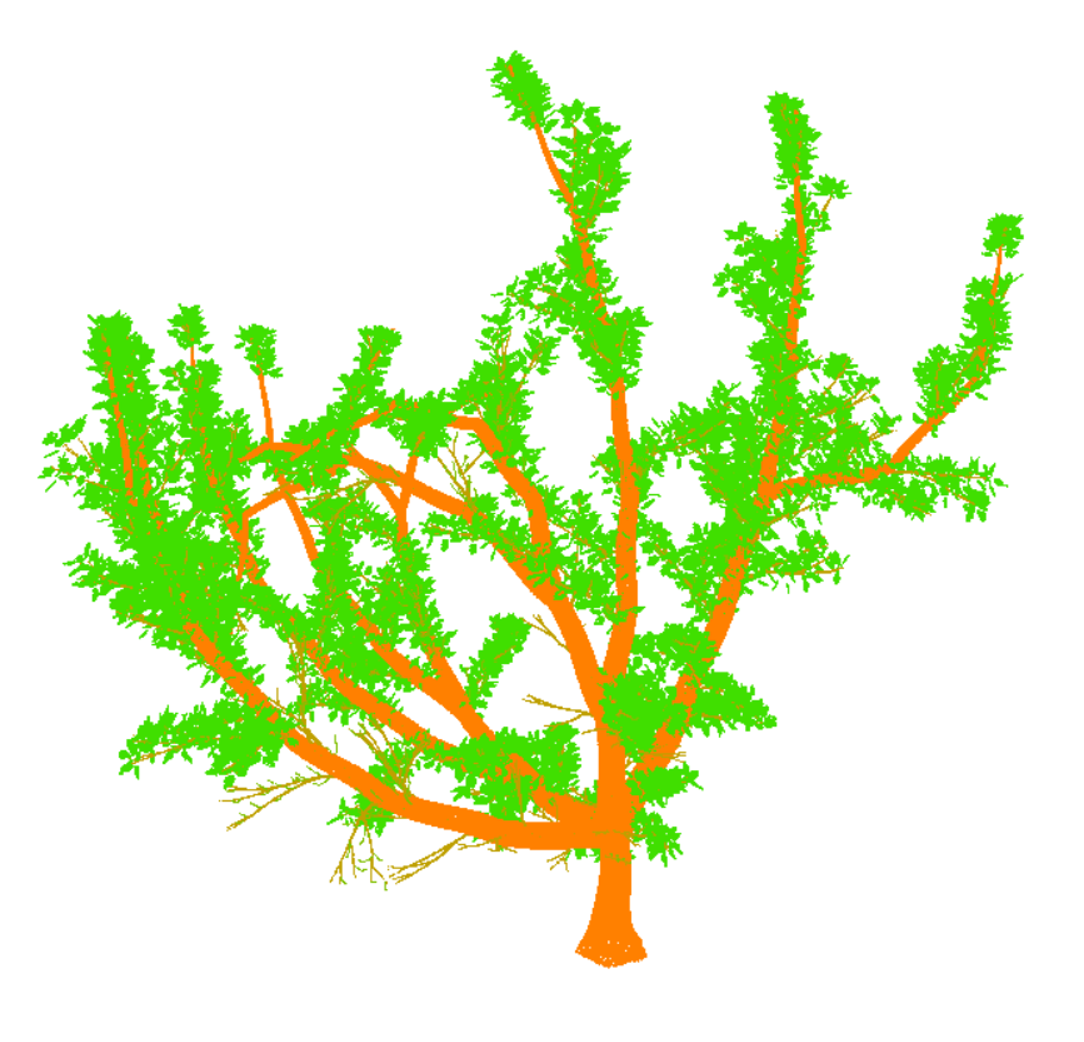}
        \caption{High density point cloud from Arbaro}
    \end{subfigure}
    ~
    \begin{subfigure}[t]{\columnwidth}
        \centering
        \includegraphics[width=0.4\textwidth]{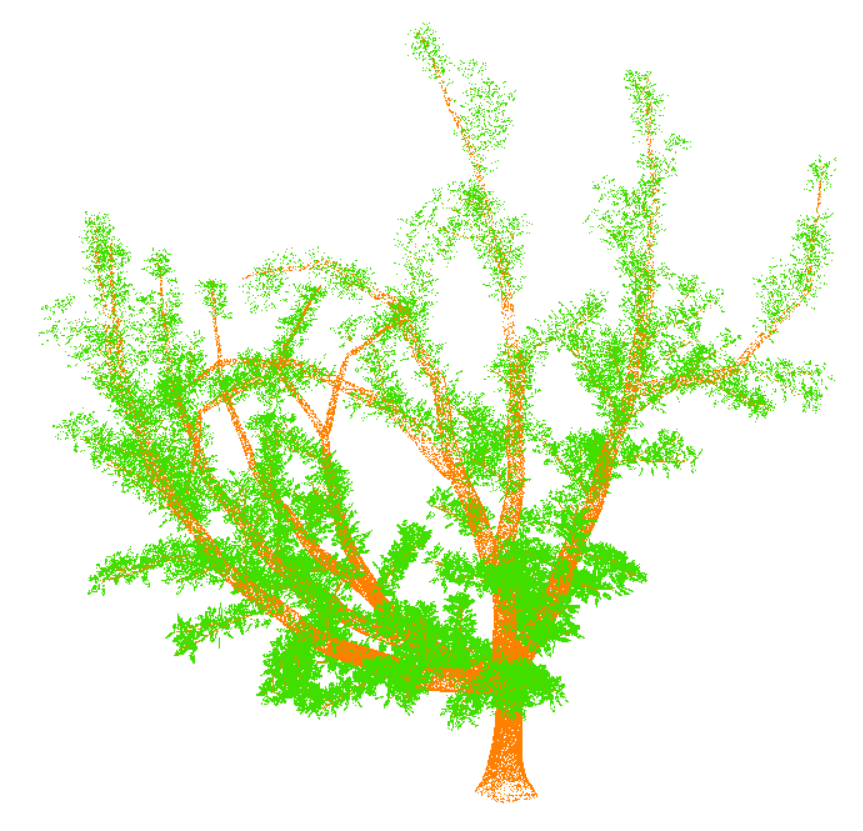}
        \caption{Simulated scan}
    \end{subfigure}
    ~
    \begin{subfigure}[t]{\columnwidth}
        \centering
        \includegraphics[width=0.4\textwidth]{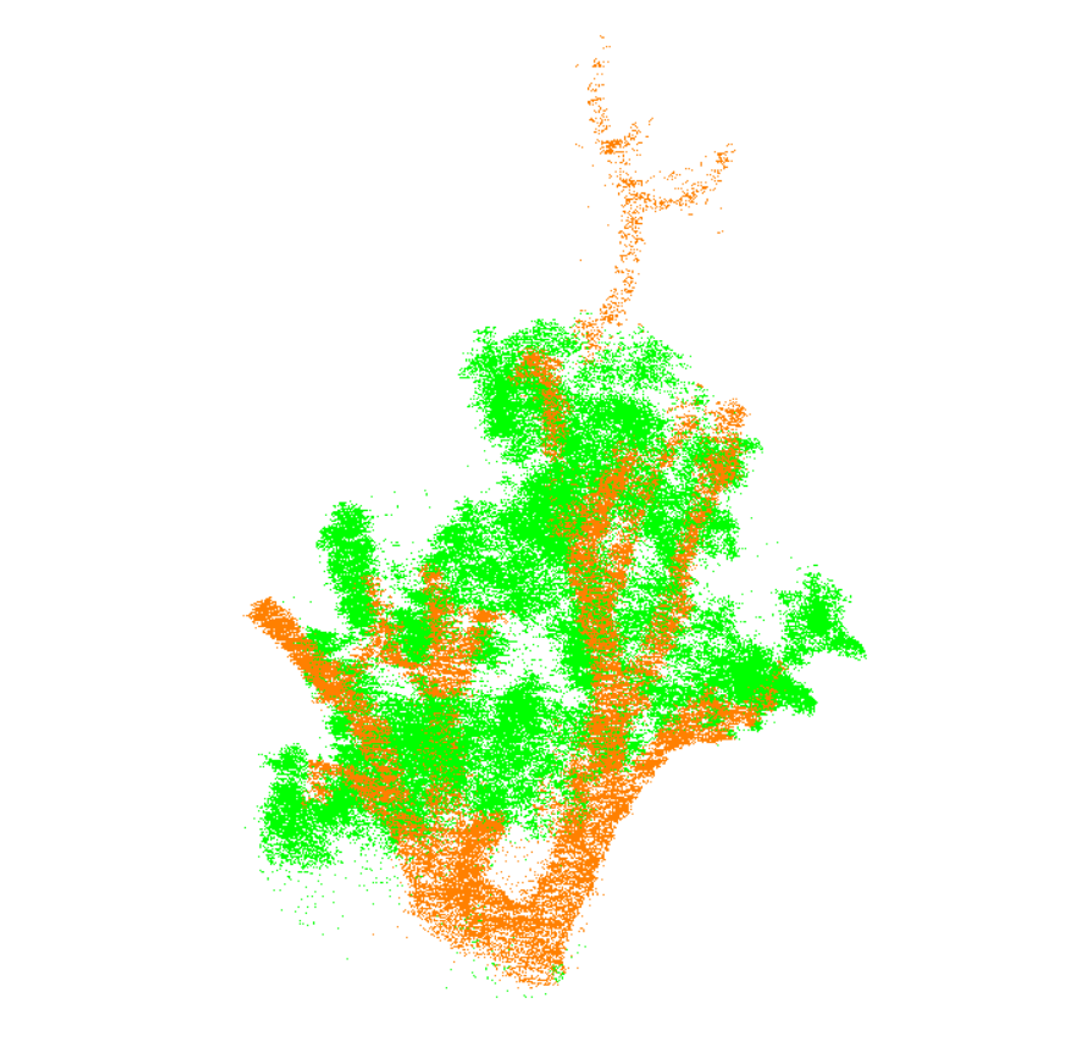}
        \caption{Real scan}
    \end{subfigure}
    \caption{Visualisation of a virtual tree before and after LiDAR simulation.  The tree shown was simulated at the focus of the Zeb1 trajectory.  The points are coloured according to their classification, from orange at Level 0 (trunk) to green at Level 3 (leaf).  A real scan of an avocado tree (with manually labelled trunks) is also shown for comparison.  Note that these trees are not the same shape originally, as the virtual tree is procedurally generated using a tree definition designed to be similar to avocado trees, rather than a model of the same tree shown in the real scan.  The similarities in the different point clouds are primarily seen in the structural characteristics, for instance the changing density of the scan towards the upper canopy.}
    \label{fig:results-individual}
\end{figure}

\begin{figure}[ht]
    \centering
    \begin{subfigure}[t]{0.5\columnwidth}
        \centering
        \includegraphics[width=\textwidth]{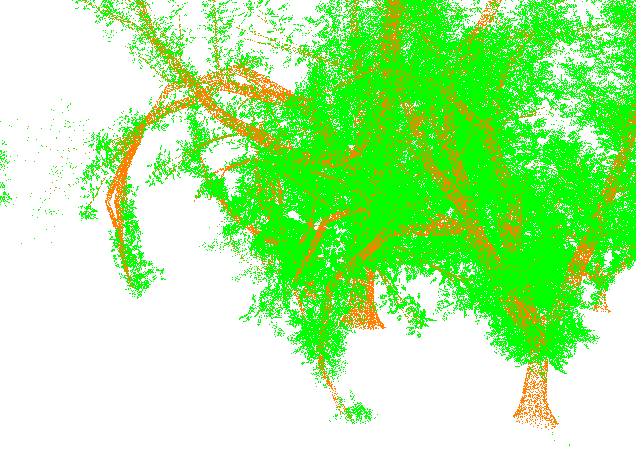}
        \caption{No noise}
    \end{subfigure}%
    ~
    \begin{subfigure}[t]{0.5\columnwidth}
        \centering
        \includegraphics[width=\textwidth]{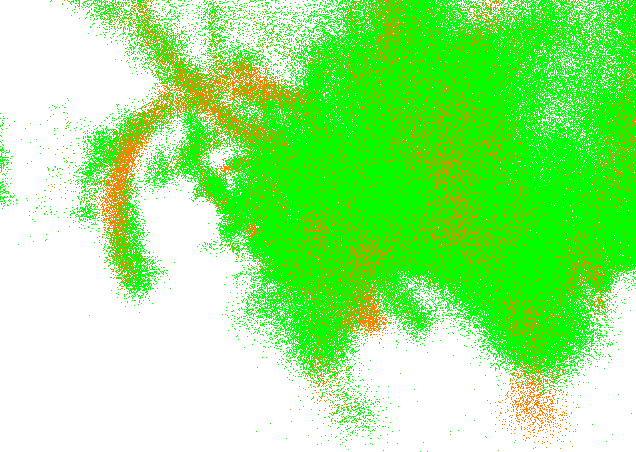}
        \caption{Noise parameter = 0.1m}
    \end{subfigure}
    ~
    \caption{Effect of simulated noise on outputs. The tree in (a) has been simulated with a Zeb1 trajectory an no added noise, while the more extreme tree in (b) has been simulated with the same trajectory and a noise level of 0.1m.  The other simulations used in this paper were simulated at 0.02m.}
    \label{fig:results-noise}
\end{figure}

We first demonstrate a visual example of the scan simulation outputs.
An example of a single simulated tree is shown in Figure~\ref{fig:results-individual}.  A close up of a tree simulated at different levels of noise is also shown in Figure~\ref{fig:results-noise}. We present an example of a real LiDAR scan using the same sensor we are simulating for comparison.

\begin{figure}[ht]
    \includegraphics[width=\columnwidth]{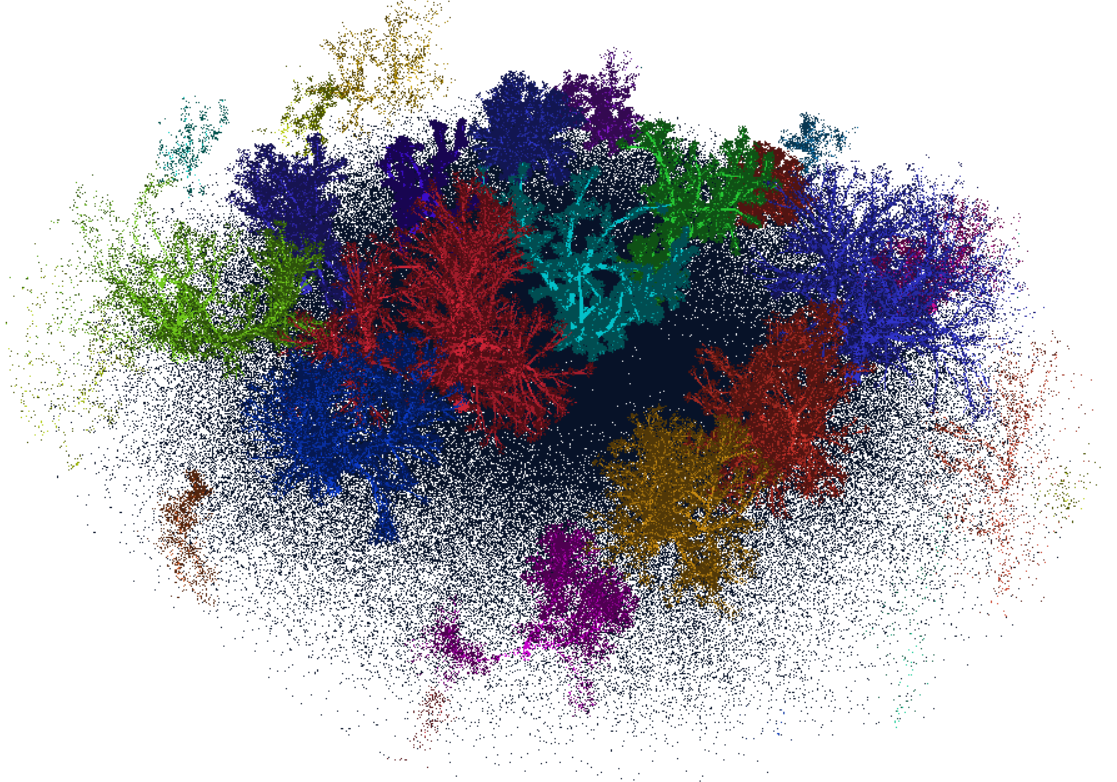}
    \caption{Visualisation of a whole stand of trees in orchard format, simulated using the Zeb1 trajectory.  Trees are coloured according to their unique ID, with brighter colours for trunk matter.  Viewing in colour is recommended.}
    \label{fig:results-stand}
\end{figure}

A visualisation of an entire scan simulated using the Zeb1 trajectory is also shown in Figure~\ref{fig:results-stand}

\subsection{Validation experiments}

Here we present results for experiments validating our method using scan density.

Table~\ref{tab:validation} presents numerical results for the datasets used in validation.  All the trees in this set are either real avocado trees or trees generated using an Arbaro definition designed to replicate avocado trees.

\begin{table}[ht]
    \begin{tabularx}{\columnwidth}{Xlllll}
                                    & \multicolumn{1}{l}{Real} & \multicolumn{1}{l}{Control} & \multicolumn{1}{l}{Simulated} \\
        Number of points            & 5467474                  & 2082868                     & 2047514                       \\
        \% of points occluded       & \multicolumn{1}{l}{N/A}  & 18.2\%                      & 56.3\%                        \\
        Average density $pts/m^{3}$ & 11518.8                  & 3267.3                      & 6813.1                        \\
        Stddev density $pts/m^{3}$  & 27152.3                  & 3155.5                      & 15872.8
    \end{tabularx}
    \caption{Experimental results for validation experiments.  All stands shown were simulated using the same sensor definition, modelled after the sensor used to generate the real scan.  The simulated cloud here was generated using "Avocado" style trees, "Zeb1" trajectory and single-plane sensor shape.}
    \label{tab:validation}
\end{table}

\begin{figure}[ht]
    \centering
    \begin{subfigure}[t]{\columnwidth}
        \centering
        \includegraphics[width=\textwidth]{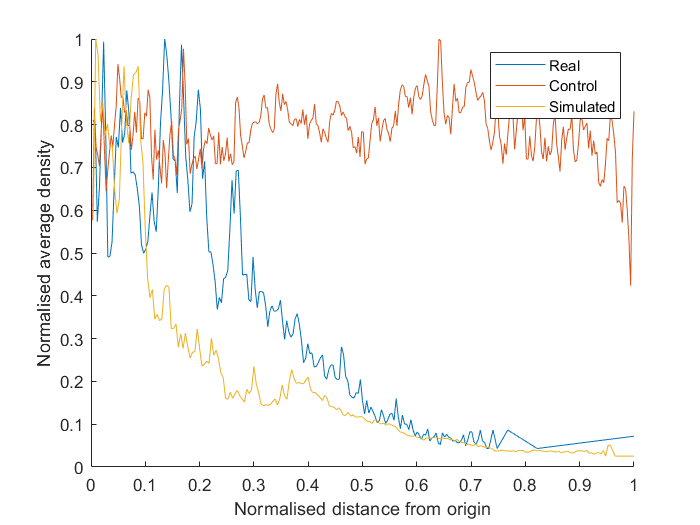}
        \caption{Density against XY distance from the origin}
    \end{subfigure}
    ~
    \begin{subfigure}[t]{\columnwidth}
        \centering
        \includegraphics[width=\textwidth]{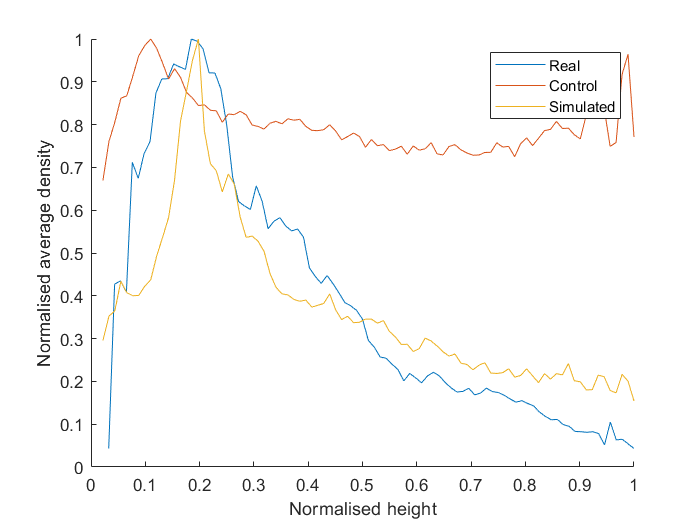}
        \caption{Density against height}
    \end{subfigure}
    \caption{Validation results for scan density.  Here we present density as a function of XY distance from the origin and the height above the ground.  The control was generated using simple mesh sampling such that the number of points was close to the simulated results. Note that the distances and density are normalised due to the difference in shape between the real and simulated trees.}
    \label{fig:results-density-graphs}
\end{figure}

We validate the scan density by plotting the scan density as a function of location and comparing the results for a real Zeb1 scan, a control point cloud generated through random sampling, and a simulated Zeb1 scan.
The scan density was considered in two contexts.  First, the density was examined as a function of distance from the origin of the point cloud.  This was chosen because the Zeb1 trajectory is specifically targeted at the central tree, leaving trees on the periphery less densely scanned as a result.
Secondly, the density as a function of height was considered.  This was because the scan density is low at the higher ends of the tree due to occlusion and beam scattering during scanning.
The plots for these functions, measured across all three point clouds, are shown in Figure~\ref{fig:results-density-graphs}.

Results from our applicability validation are shown in Table~\ref{tab:results-classeg}.
The values presented are the mean and variance for the primary metric used to quantify the performance for classification and segmentation, namely F1 score and V-Measure (\cite{rosenberg2007v}) respectively.
An unpaired t-test was also performed between the two datasets to test whether they were statistically similar, and the p value is reported.

\begin{table}[]
    \centering
    \begin{tabular}{lcc}
                         & \multicolumn{1}{l}{Classification} & \multicolumn{1}{l}{Segmentation} \\
        Real mean        & 0.4394                             & 0.9047                           \\
        Real variance    & 0.0148                             & 0.0082                           \\
        Virtual mean     & 0.5123                             & 0.8824                           \\
        Virtual variance & 0.0048                             & 0.001                            \\
        T-test p value   & 0.0163                             & 0.0422
    \end{tabular}
    \caption{Results for trunk classification and individual segmentation for real versus simulated trees (orchard, avocado).  Virtual results are over a range of tree spacings from 4m to 8m.}
    \label{tab:results-classeg}
\end{table}

\subsection{Applications}

For our applications, we process new datasets using alternative tree definitions and stand layouts.  A numerical summary of all these experiments, detailing the occlusion and density of the results, is provided in Table~\ref{tab:application}.

\begin{table}[ht]
    \begin{tabularx}{\columnwidth}{Xrrrr}
                                                   & \multicolumn{4}{l}{Avocado orchard}                                                                                       \\
                                                   & \multicolumn{1}{l}{Control}           & \multicolumn{1}{l}{Zeb1} & \multicolumn{1}{l}{Mobile} & \multicolumn{1}{l}{Drone} \\
        Number of points                           & 2082868                               & 2047514                  & 2110775                    & 563379                    \\
        \% of points occluded                      & 18.17\%                               & 56.28\%                  & 50.68\%                    & 64.26\%                   \\
        Average density (pts/m\textasciicircum{}3) & 3267.3                                & 6813.1                   & 6433.6                     & 2564.7                    \\
        Stddev density (pts/m\textasciicircum{}3)  & 3155.5                                & 15872.8                  & 13856.3                    & 2835.9                    \\
                                                   & \multicolumn{4}{l}{Aspen orchard}                                                                                         \\
        Number of points                           & 2713915                               & 1482655                  & 1447140                    & 1620340                   \\
        \% of points occluded                      & 14.42\%                               & 52.42\%                  & 45.58\%                    & 38.66\%                   \\
        Average density (pts/m\textasciicircum{}3) & 3627.5                                & 3466.3                   & 2957.6                     & 3011.2                    \\
        Stddev density (pts/m\textasciicircum{}3)  & 4021.1                                & 5600.7                   & 3765.3                     & 3111.9                    \\
                                                   & \multicolumn{4}{l}{Aspen Forest}                                                                                          \\
        Number of points                           & 3548336                               & 1428058                  & 1621777                    & 1925538                   \\
        \% of points occluded                      & 13.86\%                               & 55.74\%                  & 53.71\%                    & 48.69\%                   \\
        Average density (pts/m\textasciicircum{}3) & 3915.2                                & 3356.1                   & 3365.1                     & 3486.7                    \\
        Stddev density (pts/m\textasciicircum{}3)  & 4520.1                                & 6806.7                   & 5677.4                     & 4641.1                    \\
                                                   & \multicolumn{4}{l}{Macadamia orchard}                                                                                     \\
        Number of points                           & 3512419                               & 4973296                  & 3311894                    & 1883430                   \\
        \% of points occluded                      & 18.15\%                               & 44.07\%                  & 44.57\%                    & 48.04\%                   \\
        Average density (pts/m\textasciicircum{}3) & 3686.0                                & 7392.2                   & 4910.2                     & 2987.5                    \\
        Stddev density (pts/m\textasciicircum{}3)  & 4844.0                                & 43754.8                  & 10491.7                    & 5525.6
    \end{tabularx}
    \caption{Experimental results for applications.  All stands shown were simulated using the same sensor definition, modelled after the sensor used to generate the real scan.  These are all virtually generated trees.}
    \label{tab:application}
\end{table}

\begin{figure}
    \centering
    \begin{subfigure}[t]{0.5\columnwidth}
        \centering
        \includegraphics[width=\textwidth]{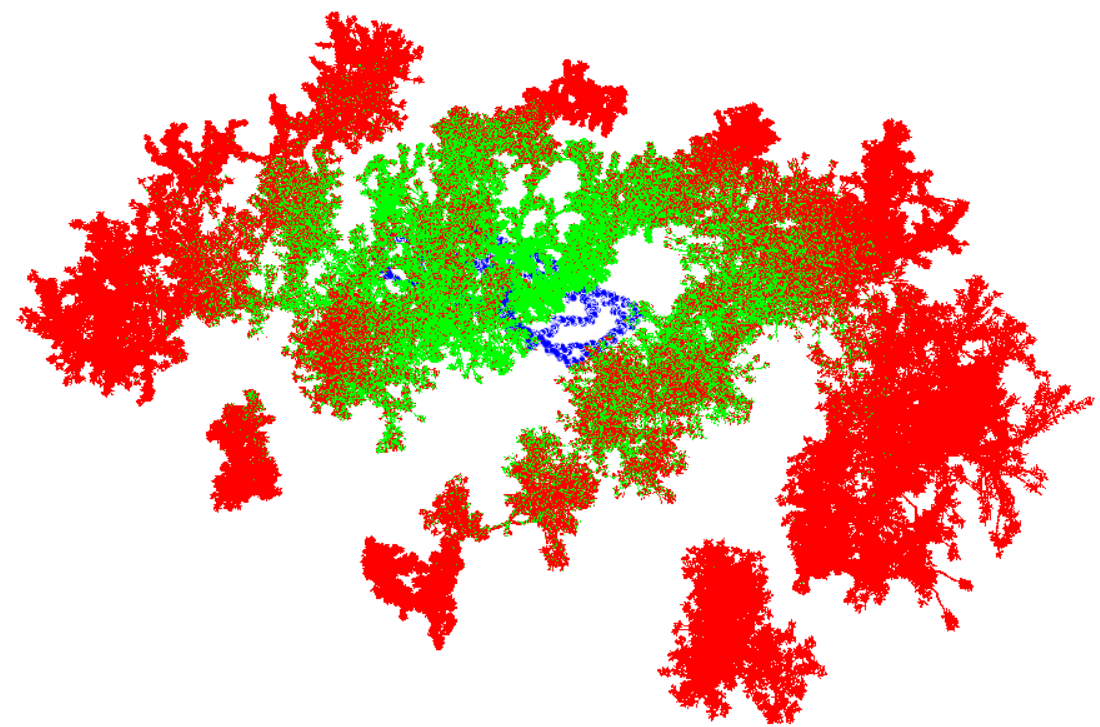}
        \caption{Zeb1}
    \end{subfigure}%
    ~
    \begin{subfigure}[t]{0.5\columnwidth}
        \centering
        \includegraphics[width=\textwidth]{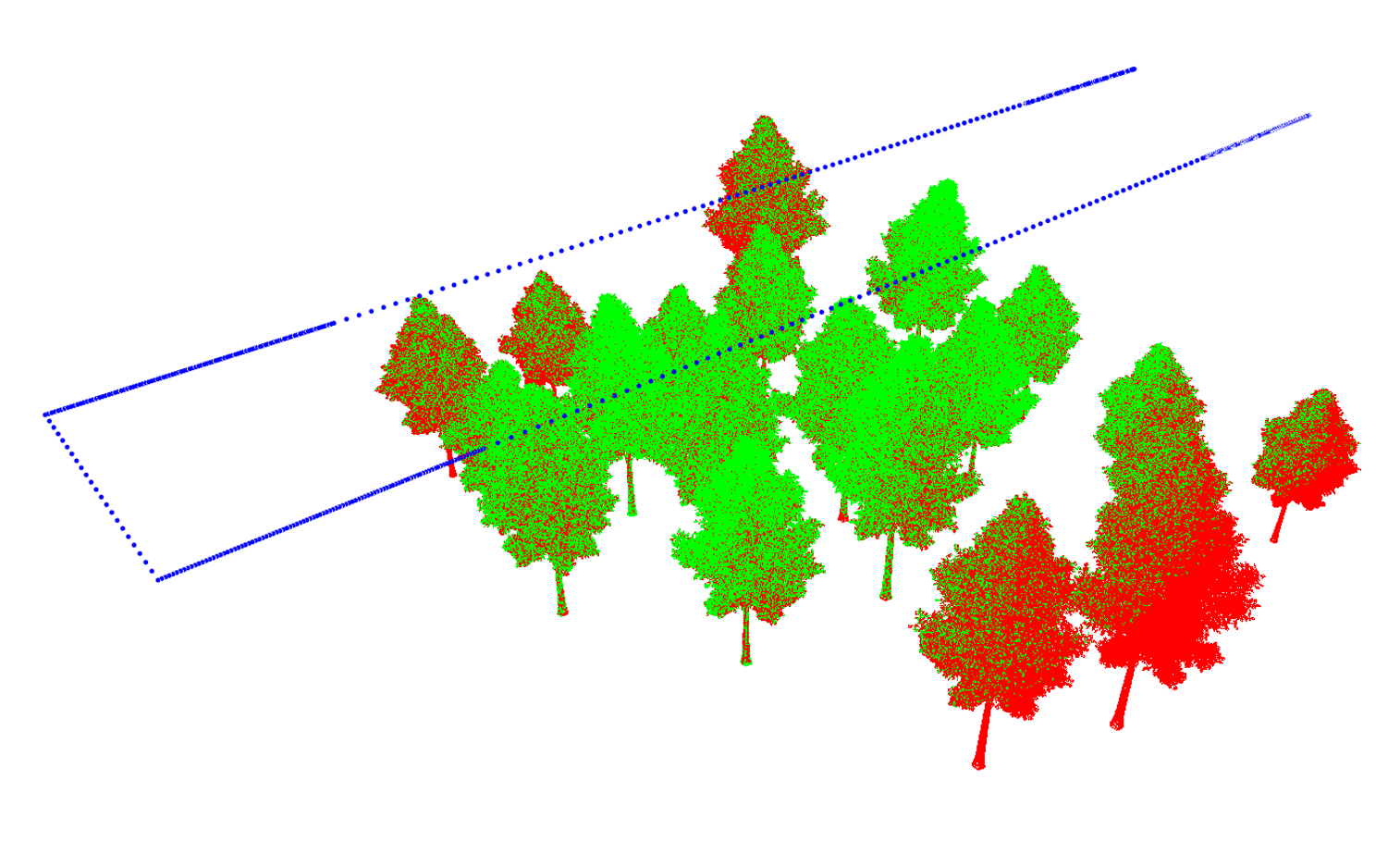}
        \caption{Forest (drone)}
    \end{subfigure}
    ~
    \begin{subfigure}[t]{0.5\columnwidth}
        \centering
        \includegraphics[width=\textwidth]{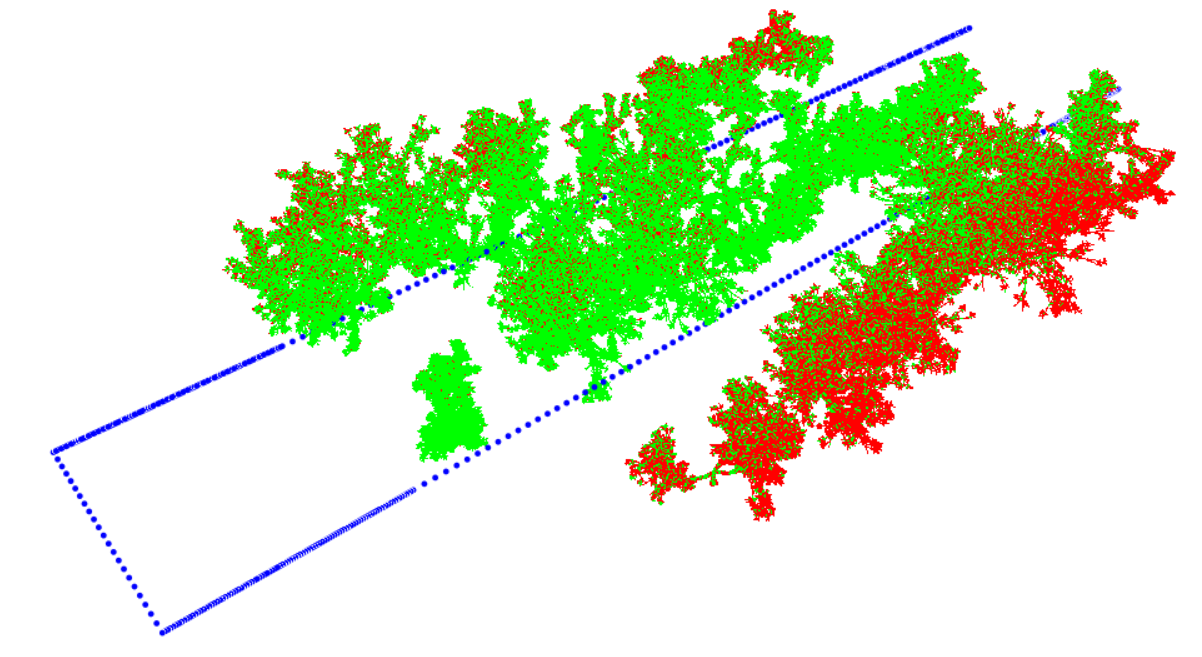}
        \caption{Mobile}
    \end{subfigure}%
    ~
    \begin{subfigure}[t]{0.5\columnwidth}
        \centering
        \includegraphics[width=\textwidth]{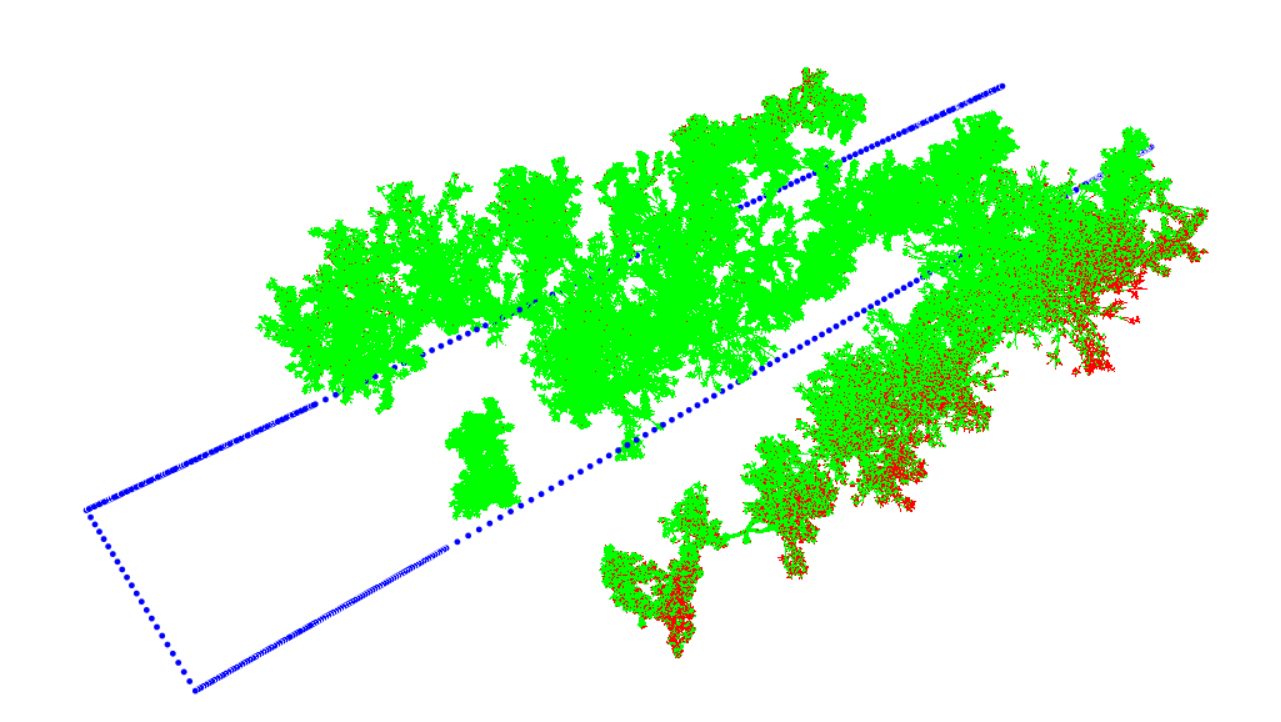}
        \caption{Mobile (9-beam LiDAR)}
    \end{subfigure}
    ~

    \caption{Examples of occlusion maps for different trajectory types as compared to the control set.  Points from the original stand which are visible in the simulated scan are in green, the others are in red.  The trajectories used to generate the scans are shown in blue where applicable.}
    \label{fig:results-occlusion-pics}
\end{figure}

For a visual example of the results, an occlusion map of each trajectory is presented in Figure~\ref{fig:results-occlusion-pics}, along with the aerial trajectory for the forest stand.  Points in red are considered occluded while points in green are considered scanned.
We also include a 9-beam LiDAR for comparison with the Mobile trajectory, demonstrating the difference in occlusion using a LiDAR with more data in the same trajectory.

\section{Discussion}
\label{sec:discussion}

In this section we discuss the results presented in the previous section, and suggest areas for future work.
Generally, the outputs of SimTreeLS present point clouds which are similar in structure and form to those generated by real LiDAR, and are easy to customise for a particular application.

The differences in real versus virtual trees can generally be characterised as a difference tree shapes and noise characteristics.
One of the main causes of discrepancies is our current inability to simulate tree motion.
Regardless of the method used for LiDAR scanning, the motion of the tree due to wind will introduce potentially significant noise, depending on the strength of the wind.
This effect can be seen in Figure~\ref{fig:results-individual}, where branches present with greater thickness than the virtual results, and leaves are shown as less distinct.

\subsection{Validation}
% What did the experiments show?
The validation results show that the outputs are reasonable approximations of reality.
Given that the simulated trees were not identical to the real ones (instead just a rough facsimile), we did not expect perfect replication, but rather a strong correlation.
The number of points shown in Table~\ref{tab:validation} suggest that the trees generated by Arbaro are potentially smaller than the real ones, leading to a smaller final point cloud.
However, we also see the relative difference between average and standard deviation of density being similar, suggesting a similar pattern of scanning, as opposed to the control.
Figure~\ref{fig:results-density-graphs} reinforces this point further, with both graphs for control being clearly different to the real ones, but the simulated graphs sharing characteristics with the real data.
This implies that the simulated scans generate data which is structurally similar to that generated by real sensors in the case of the Zeb1 trajectory.

The results of classification and segmentation of real and virtual data presented in Table~\ref{tab:results-classeg} further illustrate how SimTreeLS generates data with similar structure to real LiDAR scans.
The results of the t-test imply that the two datasets are distinct, with a p value of 0.04 for segmentation and 0.01 for classification.
The disparity between the datasets could be explained by errors in manual labelling due to indistinct areas and tree overlap for the real data, or because the virtual trees are merely approximations of the avocado tree structure.

\subsection{Applications}

% What did the experiments show?
The results for our applications illustrate the flexibility of SimTreeLS, and show that alternate trajectories and tree definitions also produce realistic results.
The additional stands tested, shown in Figure~\ref{fig:method-demos}, demonstrate particular characteristics not shown by the avocado orchard stand used in the validation experiments.
The "macadamia" tree definition is much wider than the other trees, so using the same stand definition leads to a much denser orchard, which is not unusual in commercial practices.
The "aspen" trees on the other hand have their foliage high up in the canopy, which is above the focus of the Zeb1 trajectory, and significantly denser than the other trees, which should show a high occlusion.
The forest stand is interesting since the trajectories were designed to optimise scanning of the orchard definition, with the Zeb1 trajectory specifically focused on the central tree.
This suggests the ability for SimTreeLS to be used in evaluation and optimisation of trajectories for particular contexts.  For instance, given a known sensor in a forestry setting and a set of constraints on potential trajectories, one could be found which maximised scan density or minimised occlusions.

The sensor definition can also make a big difference, for instance the mobile trajectory shown in Figure ~\ref{fig:results-occlusion-pics} demonstrates the occlusion with a 9-beam sensor is significantly lower than one with a single beam.
It could be anticipated that more data should be detected with the larger sensor, but the additional angles provided significantly more

% What do the results mean?
The results presented in Table~\ref{tab:application} support the intuitive implications of these different stands, as well as suggesting useful insights.
For the macadamia trees, the high density of the foliage leads to a large standard deviation across the different trajectories, since the outside of the tree is well scanned and the inside is significantly more occluded.
The aspen trees are the only examples where the aerial trajectory has the largest number of points in the output as well as the smallest occlusion.  The standard deviation also suggests a very even distribution of scanned points, whereas the Zeb1 and Mobile trajectories miss a lot of the detail in the trees.
This is further exasperated in the forest layout, where the mobile trajectory in particular (which only catches the foliage in the edges of the scanner) misses significant amounts of the trees since it does not drive along the rows of trees here.

\subsection{Future work}

SimTreeLS shows promising results in procedural dataset generation, which has the potential to be helpful in deep learning applications using transfer learning.
Future work should include validating this proposal so as to make simulated data more appealing for applications where data capture or manual labelling is difficult prohibitive.

The system could also easily be extended to include items beyond trees.  Significant amounts of LiDAR research is developed on sampled meshes, similar to the "control" used in our experiments, but SimTreeLS could use any sensor definition and trajectory to scan arbitrary meshes.  For example, there are applications in the construction industry using LiDAR, and there is often a digital model of the design which could be scanned in this way.

With regards to the system itself, many modern LiDAR sensors are able to capture colour or intensity information, as well as multiple returns for each beam.
These features would be a simple inclusion in the pipeline used to generate results.

Other potentially exciting extensions of SimTreeLS include the option to suggest an optimal trajectory given a stand and sensor definition.  This may be able to inform users as to which approach to take given a particular context, and could be used to minimise time required for fieldwork.

\section{Conclusion}
\label{sec:conclusion}
We presented a system, SimTreeLS, for generating simulated LiDAR scans of procedurally generated trees in agricultural and forestry contexts.  Validation experiments have shown that the generated data is similar in nature to real LiDAR scans, and several capabilities have been explored and visualised.

\subsubsection*{Acknowledgements}
This work is supported by the Australian Centre for Field Robotics (ACFR) at The University of Sydney. For more information about robots and systems for agriculture at the ACFR, please visit http://sydney.edu.au/acfr/agriculture.

\bibliography{references}{}
\end{document}